# 1/f fluctuations of DNA temperature at thermal denaturation

Marc JOYEUX[(#)], Sahin BUYUKDAGLI and Michaël SANREY

*Laboratoire de Spectrométrie Physique (CNRS UMR 5588),*

*Université Joseph Fourier - Grenoble 1,*

*BP 87, 38402 St Martin d'Hères, FRANCE*

**Abstract** - We theoretically investigated the temperature fluctuations of DNA close to denaturation and observed a strong enhancement of these fluctuations at the critical temperature. Although in a much lower frequency range, such a sharp increase was also reported in the recent experimental work of Nagapriya *et al* [Phys. Rev. Lett. 96, 038102 (2006)]. We showed that there is instead no enhancement of temperature fluctuations when the dissipation coefficient γ in Langevin equations is assumed to be larger than a few tens of $ps^{-1}$, and pointed out the possible role of the solvent in real experiments. We sought for a possible correlation between the growth of large bubbles and the enhancement of temperature fluctuations but found no direct evidence thereof. Finally, we showed that neither the enhancement of fluctuations nor the $1/f$ dependence are observed at the scale of a single base pair, while these properties show up when summing the contributions of a large number of base pairs. We therefore conclude that both effects result from collective motions that are facilitated by the divergence of the correlation length at denaturation.

[(#)] email : Marc.JOYEUX@ujf-grenoble.fr

# I - Introduction

The physics of DNA thermal denaturation (or melting), that is, the separation of the two strands upon heating, is a subject which has a long history [1-3] because it can be viewed as a preliminary for understanding transcription and/or replication. The wide interest in this problem is also motivated by the fact that UV absorption spectroscopy experiments indicate that the whole denaturation process looks like a chain of sharp first-order like phase transitions, in the sense that large portions of inhomogeneous DNA sequences separate over narrow temperature intervals [4]. Statistical models ascribe the discontinuous character of the transitions to excluded volume effects [5,6], while dynamical ones invoke instead a strong entropic effect due to significant softening of the DNA chain at large displacements [7,8]. In a recent Letter [9], Nagapriya, Raychaudhuri and Chatterji reported the measurement of the fluctuations of DNA temperature close to the critical temperature, an issue which had apparently not been addressed before, neither experimentally nor theoretically. They observed a strong enhancement of the low-frequency fluctuations (0.01 to 1 Hz range) at the melting temperature and suggested that this might be the result of coexisting phases that are in dynamical equilibrium. Triggered by these experimental findings, we performed molecular dynamics calculations to investigate this point theoretically. Although in a much higher frequency range (1 MHz to 10 GHz), we also observed a strong enhancement of temperature fluctuations at the melting temperature. Calculations additionally showed that the power spectra of temperature fluctuations follow an unexpected $1/f$ law over several decades. The purpose of this paper is to report on our results and to discuss the possible origin of the $1/f$ fluctuations.

The remainder of this paper is organized as follows. Computation procedures are briefly sketched in Section II. The enhancement of temperature fluctuations and the $1/f$ dependence are put in evidence in Section III on the basis of molecular dynamics simulations performed with three rather different mesoscopic models. The extent to which these results depend on the value of the dissipation coefficient γ in Langevin equations is investigated in Section IV. In Section V, we discuss the lack of correlation between the growth of large bubbles and the enhancement of temperature fluctuations. Finally, we study in Section VI how the enhancement of fluctuations and the $1/f$ dependence arise from the summation of the contribution of an increasing number of base pairs and discuss these properties in terms of slow collective motions and the divergence of the correlation length at denaturation.

**II - Models and simulations**

Results presented in this paper were obtained with three different dynamical models that were shown to reproduce the denaturation dynamics of homogeneous and/or inhomogeneous DNA sequences quite accurately. All of them are of the form

$$E = E_K(\dot{\mathbf{q}}) + E_P(\mathbf{q}) , \qquad (\text{II-1})$$

where $E_K(\dot{\mathbf{q}})$ and $E_P(\mathbf{q})$ are the kinetic and potential energies, which are expressed as a function of $n$ position coordinates $\mathbf{q} = (q_1, q_2, ..., q_n)$ and their derivatives with respect to time $\dot{\mathbf{q}} = (\dot{q}_1, \dot{q}_2, ..., \dot{q}_n)$. Since for the three models the kinetic energy is taken in the form of a diagonal quadratic expansion

$$E_K(\dot{\mathbf{q}}) = \sum_{k=1}^{n} \alpha_k \dot{q}_k^2 , \qquad (\text{II-2})$$

where the $\alpha_k$ denote positive real parameters, the temperature of the sequences is straightforwardly estimated from

$$T = \frac{2 E_K(\dot{\mathbf{q}})}{n \, k_B} . \qquad (\text{II-3})$$

The three models also agree in describing the interaction between two paired bases with a Morse potential, as suggested by Prohofsky *et al* [10]. The models however differ by the number of degrees of freedom which are used to describe each base pair (from 1 to 4) as well as the form of the stacking energy, *i.e.* the interaction between two successive base pairs :

In the Dauxois-Peyrard-Bishop (DPB) model, each base pair $k$ is described by a single variable $q_k \equiv y_k$, which represents the transverse stretching of the hydrogen bonds connecting the two bases, so that the number $n$ of degrees of freedom is equal to the length $N$ of the sequence. Moreover, the stacking energy is modelled by an harmonic potential plus an exponential correction that reduces the stacking interaction at large inter-strand separations and makes the denaturation much sharper. See Eqs. (1)-(2) of Ref. [7] for the explicit expression of the DPB model.

The second (JB) model was derived in our group to take into account the fact that stacking energies are necessarily finite. Each nucleotide is modelled by a point mass which moves along the axis connecting the two strands, so that base pair $k$ is described by two position coordinates $u_k$ and $v_k$ (for this model $n$ is equal to $2N$). Stacking interactions are represented by gaussian holes, whose depths are the stacking enthalpies determined from thermodynamics calculations [11]. See Eq. (4) of Ref. [8] for the explicit expression of the JB model.

In the last, more complex BSJ model, stacking interactions are again represented by gaussian holes, but the bases and the associated phosphate/sugar groups are modelled by separate sets of point masses. The phosphate/sugar groups move along the axis connecting the two strands while the bases rotate in the plane perpendicular to the sequence axis. The motion of each base pair is thus described by two position coordinates $x_k$ and $X_k$ plus two angles $\gamma_k$ and $\Gamma_k$ ($n = 4N$). See Eq. (1) of Ref. [12] for the explicit expression of the BSJ model.

Note that the JB and BSJ models were shown to give results in good agreement with statistical ones for inhomogeneous sequences [8,12]. Molecular dynamics simulations consisted in integrating numerically, with a second-order Brünger-Brooks-Karplus integrator [13] and time steps of 10 fs, the equations of motion

$$\alpha_k \frac{d^2 q_k}{dt^2} = -\frac{\partial E_P}{\partial q_k} - \alpha_k \gamma \frac{dq_k}{dt} + w(t)\sqrt{2\alpha_k \gamma k_B T} ,  \qquad \text{(II-4)}$$

where $\gamma$ is the dissipation coefficient and $w(t)$ a normally distributed random function with zero mean and unit variance. The second and third terms in the right-hand side of Eq. (II-4) model the dissipation and stochastic collisions exerted by the solvent on DNA. These terms are crucial for heating the sequence to the requested temperature, since this is achieved by slowly increasing the $T$ variable, standing for the temperature of the solvent, which appears in the collision term. A dissipation coefficient $\gamma = 5$ ns$^{-1}$ was assumed for the heating phase. At the end of this phase, the time average of the temperature of the DNA sequence, obtained from Eq. II-3, must be equal to the temperature imposed by the solvent in Eq. (II-4). Once the correct temperature is reached, the last two terms in Eq. (II-4) can be either maintained (Langevin dynamics) or dropped (Hamilton dynamics). For each set of initial conditions, spectral power density (SPD) was finally computed according to

$$P(f) = \frac{2}{t_{max}} \left| \int_0^{t_{max}} (T - \bar{T}) \exp(-2\pi i f t) dt \right|^2 , \qquad \text{(II-5)}$$

where $T$ is the temperature of the DNA sequence, $t_{max}$ denotes the integration time and $\bar{T}$ the average of $T$ between 0 and $t_{max}$. Practically, we set $t_{max}$ to 0.5 μs and systematically averaged $P(f)$ over 10 to 20 different sets of initial conditions at the same temperature. Finally, we slightly smoothed the obtained curves to further reduce the speckle noise.

**III - Enhancement of temperature fluctuations at thermal denaturation**

Figs. 1 to 3 show SPD curves obtained from Hamilton dynamics calculations for, respectively, a 2000 base pair (bp) homogeneous sequence and the DPB model [7], a 2399 bp inhibitor of the hepatocyte growth factor activator [14] and the JB model [8], and the 1793 bp human β-actin cDNA (NCB entry code NM_001101) and the BSJ model [12]. All computations were performed with open boundary conditions. The melting temperatures for these three systems, *i.e.* the temperature at which 50% of the base pairs are open, are about 354 K, 335 K and 385 K, respectively. These temperatures were determined from melting curves calculated as in Fig. 3 of Ref. [8].

Although in a much higher frequency range, these curves clearly exhibit the same behaviour as the experimental results of Ref. [9] : (i) fluctuations vary slowly with temperature below denaturation, (ii) they increase strongly at the critical temperature, especially in the red part of the spectrum, (iii) they become again smaller as soon as the two strands are fully separated. Most importantly, comparison of Figs. 1 to 3 indicates that this behaviour is not model-dependent in the sense that it is exhibited by three quite different models. It can therefore reasonably be speculated that the enhancement of fluctuations is characteristic of DNA melting and not of the peculiar model used to describe DNA melting.

All the sequences we tested display the characteristic behaviour of Figs. 1 to 3, whatever the sequence and the model. This includes homogenous and regular sequences as well, whereas Nagapriya *et al* noted that "fluctuation is larger in heterogeneous samples" [9]. However, it appears that this conclusion is based on the properties of a single short 100 bp regular ATAT... sequence, while all the sequences we investigated contained more than 1500 bp, essentially because it is numerically rather difficult to impose a given temperature to a short sequence. Since the dynamics of short sequences is particularly sensitive to the opening at the two ends, it may differ qualitatively from that of longer sequences. It might therefore be interesting to check whether measurements performed on longer regular sequences confirm the conclusion of Nagapriya *et al* or match the calculations in the higher frequency range.

Another feature, which is common to all the sequences and the three investigated models, is the fact that SPD varies as $1/f$ over several decades. We will come back to this point in Sect. VI.

**IV - Influence of the dissipation coefficient γ on temperature fluctuations**

As we already mentioned, the results presented in Sect. III were obtained by integrating Hamilton's equations of motion, which means that γ was turned to zero in Eq. (II-4) after the heating phase. Depending on the value of the dissipation coefficient γ, quite different results may however be obtained when integrating Langevin's equations of motion instead of Hamilton's ones. This is due to the fact that, in the thermodynamic limit of infinitely long sequences, the averages of thermodynamic observables in the micro-canonical (Hamiltonian dynamics) and canonical (Langevin dynamics) ensembles do coincide, but not their fluctuations [15]. For example, Fig. 4 shows SPD curves for the 2399 bp inhibitor at different temperatures obtained by integrating Langevin's equations of motion for the JB model [8]. We used a dissipation coefficient γ=5 ns$^{-1}$ that lies in the range of commonly accepted values for this parameter. In contrast with Fig. 2, the curves in Fig. 4 do not exhibit significant changes at the melting temperature. To characterize the influence of γ on temperature fluctuations in more detail, we further plotted in Fig. 5 the SPD curves at the melting temperature (335 K) for increasing values of γ. It is seen that values of γ up to 0.05 ns$^{-1}$ have little influence on temperature fluctuations down to the MHz range. In contrast, dissipation coefficients of 0.5 ns$^{-1}$ and 5 ns$^{-1}$ damp all oscillations with frequencies smaller than 10 MHz and 100 MHz, respectively, while the SPD curve obtained with $\gamma=50$ ns$^{-1}$ departs from those obtained with lower values of $\gamma$ in the whole $10^6$-$10^{10}$ Hz range.

Conclusion therefore is that erratic collisions with solvent molecules may well damp the strong enhancement of temperature fluctuations experienced by isolated DNA sequences. Whether this happens in the range of "physical" values of $\gamma$ or not remains at present an open question. Nonetheless, we noted with interest that the measurements reported in Ref. [9] were performed after the solution was allowed to evaporate off, in order that the thermal mass of the solvent does not reduce experimental sensitivity. Of course, most DNA sequences are still in contact with each other, with residual solvent molecules and with the thermometer, but they are probably more isolated than in solution, which may result in a lower effective value of the dissipation coefficient $\gamma$. To fix this question regarding the possible effect of the solvent, it might be interesting - if technically feasible - to repeat the measurements for increasing quantities of solvent and/or for solvents which interact more or less strongly with DNA.

## V - Temperature fluctuations *versus* bubbles dynamics

It is well known that large bubbles develop precisely in the temperature range where enhancement of fluctuations is observed [3,8,12]. It is thus legitimate to wonder whether there exists a connection between the enhancement of temperature fluctuations and bubble dynamics. More precisely, the question that naturally arises is whether temperature fluctuations are larger in bubbles than in closed portions of the sequence. If this were the case, then bubble formation would be directly responsible for the observed enhancement of fluctuations at the critical temperature. In order to check this hypothesis, we investigated temperature fluctuations not only of the whole sequence but also of smaller subsections. This is illustrated in the upper plot of Fig. 6, which shows the melting profile of the 2399 bp inhibitor [14] at 325 K and 330 K. Larger values of the average distance between paired bases indicate regions where bubbles are preferentially formed, that is, regions with a higher local percentage of AT base pairs [8,12] (see the lowest plot of Fig. 6). We considered the 2399 bp sequence to consist of 11 different subsections labelled A to K. Some of these subsections (like A, D, F, H and K) are often melted at 325 K while other ones (like B, E and I) remain essentially closed at this temperature. At the melting temperature (335 K), all subsections, except for B, are open most of the time. We next plotted the SPD of temperature fluctuations for each subsection $j$ separately. For example, Fig. 7 shows, for the 11 subsections of the inhibitor sequence at 335 K, log-log plots of the product of the SPD (of each subsection) times the length $N_j$ of each subsection. It is seen that all the curves are superposed, as is also the case at 325 K. Thus, at both temperatures, (i) power spectra scale as $1/N_j$, (ii) except for this global scaling, SPD is identical for all subsections and identical to that of the complete sequence. This is clearly seen by comparing Fig. 7 with the curve for the whole sequence at 335 K in Fig. 2. In particular, the curve for subsection B (thick line in Fig. 7), which is the only closed one at 335 K, does not differ from that of open subsections.

Down to the MHz range, the enhancement of temperature fluctuations at the melting temperature is thus uniform in the chain and not localized in open subsections, *i.e.* in bubbles. At that point, it should be reminded that bubbles and closed subsections are the two phases that coexist at the critical temperature at the level of the secondary structure of DNA. As an interpretation of their experimental work in the 0.01-1 Hz range, Nagapriya *et al* suggested that the observed large fluctuations may be "a result of coexisting phases that are in dynamical equilibrium" [9]. Of course, the "coexisting phases" mentioned by these authors may well relate to the tertiary structure (or higher order conformations) of DNA. On the other hand, if they indeed relate to its secondary structure, that is to bubbles and open sections, then the calculations presented above show that the enhancement of temperature fluctuations at melting temperature is due to a mechanism which is more complex than the existence of larger fluctuations in the phase that becomes predominant at denaturation (bubbles) compared to the phase that prevails at lower temperatures (closed subsections).

Before concluding this section, it should be mentioned that power spectra disregard relative phases, so that the fact the enhancement of temperature fluctuations at the melting temperature (estimated from the increase of PSD curves) is uniform in the chain (at least down to the MHz range) does not imply that fluctuations themselves are uniform in the sequence. This can be checked in Fig. 8, which shows the time evolution of the temperature of the 2399 bp inhibitor [14] at 325 K and 335 K. In these plots, each pixel corresponds to the temperature averaged over 10 successive base pairs and 50 ns. If the diagram at 325 K (top plot of Fig. 8) and the profile shown in Fig. 6 seem to be largely uncorrelated, this is no longer the case at 335 K (bottom plot of Fig. 8). At the denaturation temperature, one indeed observes a rather clear difference between the motions of base pairs located below and above $k \approx 750$ : the former ones, which are still quite often closed, display fluctuations which look like those at 325 K, while the latter ones display fluctuations involving the coherent motion of a larger number of base pairs over longer time scales.

**VI - Collective phenomena, enhancement of fluctuations and $1/f$ dependence**

As mentioned in Sect. III, the two features that are common to all models and all sequences are (i) the enhancement of fluctuations at the melting temperature, and (ii) the fact that SPD varies as $1/f$ over several decades. For the simpler DPB and JB models [7,8], which involve only stretch degrees of freedom, the frequency range where the $1/f$ dependence holds broadens sharply at the threshold. This range extends down to 100 MHz for the DPB model at 354 K (see Fig. 1) and to less than 2 MHz for the 2399 bp inhibitor [14] at 335 K and the JB model (see Fig. 2). For the more complex BSJ model [12], which involves both stretch and bend coordinates, the frequency range where the $1/f$ dependence is observed seems instead to vary only slightly with temperature and extends down to about 100 MHz for the 1793 bp actin (Fig. 3). $1/f$ fluctuations, occasionally called "flicker noise" or "pink noise" (because located between the constant "white noise" and the $1/f^2$ "red noise"), have been observed in many areas of electronics, physics, biology, geophysics, traffic flow, financial data, *etc*, and the search for an explanation that would account for this ubiquity is one of the oldest unsolved problems in physics. If such an all-encompassing theory is still missing, the observed $1/f$ fluctuations have however in most cases been explained by *ad hoc* models. In deterministic dynamical systems, the explanation which prevails [16] is that this phenomenon takes place at the stochastic transition, that is in the energy range where quasi-regular portions of the phase space are separated from completely chaotic ones by a hierarchy of nested broken tori (also called "cantori"). $1/f$ fluctuations then appear as a consequence of the trapping of the system in the nested cantori and its subsequent release at variable rates [16]. The validity of this explanation is difficult to check in systems with many degrees of freedom, because it is no longer possible to draw Poincaré surfaces of section like those shown in Figs. 2-4 of Ref. [16]. Moreover, we checked that for all models the largest Lyapunov exponent drops abruptly to almost zero when both strands separate but that it does not display any other significant evolution at and slightly below the denaturation temperature (see for example Fig. 3 of Ref. [17]). The purpose of the remainder of this section is instead to stress that, for the ADN sequences investigated in this work, both the $1/f$ dependence and the enhancement of temperature fluctuations at the denaturation threshold are collective phenomena, in the sense that they arise from the coherence of many individual motions.

It was shown in the previous section and Fig. 7 that SPD at a given temperature scales as $1/N_j$. However, this is true only for values of $N_j$ larger than a certain threshold, which is of the order of 100-200 for the 2399 bp inhibitor [14] at 335 K and the JB model [8]. The evolution of the SPD for a broader range of $N_j$ values ($1 \leq N_j \leq 2399$) is shown in Fig. 9. It is seen that for small values of $N_j$ fluctuations scale only as $1/\sqrt{f}$ (slope -1/2) instead of $1/f$ (slope -1). As $N_j$ increases, the amplitude of higher frequency fluctuations decreases however more rapidly than that of lower frequency ones, till the $1/f$ dependence is reached at the threshold value of 100-200. For values of $N_j$ larger than the threshold, higher and lower frequency fluctuations decrease at the same rate, so that one observes the $1/f$ dependence when averaging over the whole sequence.

This point is illustrated in more detail in Fig. 10, which shows the evolution of the power of temperature fluctuations as a function of $N_j$ for exponentially increasing frequencies $f$ (2, 20 and 200 MHz and 2 GHz) and three typical temperatures : 315 K (double stranded DNA, top plot), 335 K (melting temperature, middle plot) and 345 K (single stranded DNA, bottom plot). This figure can be understood by invoking the Wiener-Khinchin theorem, which states that the SPD of a signal is equal to the Fourier transform of its autocorrelation function. This means that the SPD of temperature fluctuations computed over subsections of length $N_j$ is proportional to the Fourier transform of

$$C(N_j, \tau) = \frac{1}{N_j^2} \sum_{k=1}^{N_j} \left\langle p_k^2(t) p_k^2(t+\tau) \right\rangle + \frac{1}{N_j^2} \sum_{k=1}^{N_j} \sum_{\substack{m=1 \\ m \neq k}}^{N_j} \left\langle p_k^2(t) p_m^2(t+\tau) \right\rangle . \qquad (\text{VI-1})$$

The second term in the right-hand side of Eq. (VI-1) describes cross-correlations of fluctuations from different base pairs and vanishes if the fluctuations are independent. In the absence of correlations, the SPD is thus expected to vary as the inverse of $N_j$ if, as is the case here, the fluctuations of $p_k^2(t)$ are uniform in the sequence, *i.e.* they do not depend on the base pair position *k*. On the other hand, if all the base pairs belonging to the same subsection move as a rigid body, then the second term in Eq. (VI-1) is equal to $N_j - 1$ times the first one, so that the SPD no longer depends on $N_j$. Finally, destructive correlation (like the flip-flop of water molecules) must necessarily take place for the SPD to vary more sharply than $1/N_j$.

Examination of Fig. 10 indicates that, at the level of a single base pair, temperature fluctuations *decrease* with increasing temperatures. The *enhancement* of temperature fluctuations which is observed at the melting temperature is therefore a direct consequence of the correlations that grow up at this temperature. More precisely, it is seen that the intensity of high frequency fluctuations ($f \geq 200$ MHz) varies at all temperatures as $1/N_j$ for $N_j \geq 20$, while the slope is smaller for $N_j < 20$. This suggests that these high frequency fluctuations arise from rapid motions which however do not correlate beyond the first few neighbours. In contrast, at both 315 K and 345 K the intensity of low frequency fluctuations ($f \leq 20$ MHz) varies much more sharply than $1/N_j$ for large values of $N_j$. This indicates that excitations involving up to several hundreds of base pairs travel along the chain, so that the temperature of the complete sequence varies much less than that of given subsections. This can be checked in the top plot of Fig. 8, where diffuse darker regions with lower temperature are seen to wander almost erratically in the sequence. This phenomenon is similar to the flip-flop that is observed for neighbouring water molecules. Finally, at the melting temperature (335 K), the intensity of low frequency fluctuations ($f = 2$ MHz) varies more slowly than $1/N_j$. This is probably due to the fact that the JB model [8], as well as the DBP one [7], exhibit a well characterized first order phase transition at denaturation [18,19]. The correlation length becomes infinite at the critical temperature, thus allowing for very slow collective motions. It may therefore reasonably be hypothesized that the divergence of the correlation length at the melting temperature is responsible for the sharp extension to lower frequencies of the range where the $1/f$ dependence holds (see Figs. 1 and 2). For the more complex BSJ model [12], the frequency range where the $1/f$ dependence is observed seems instead to vary only slightly with temperature (see Fig. 3). This might indicate that denaturation is sharper in this latter model than in the two former ones and that the temperature range where the correlation length diverges is too narrow to be investigated by means of Langevin dynamics. Further work is needed to ascertain this point.

**VII - Conclusion**

To summarize, we theoretically investigated the temperature fluctuations of DNA close to denaturation and observed a strong enhancement of these fluctuations at the critical temperature. Although in a much lower frequency range, such a sharp increase was also reported in the recent experimental work of Nagapriya *et al* [9]. We showed that there is instead no enhancement of temperature fluctuations when the dissipation coefficient γ in Langevin equations is assumed to be larger than a few tens of $ps^{-1}$, and pointed out the possible role of the solvent in real experiments. We sought for a possible correlation between the growth of large bubbles and the enhancement of temperature fluctuations but found no direct evidence thereof. Finally, we showed that neither the enhancement of fluctuations nor the $1/f$ dependence are observed at the scale of a single base pair, while these properties show up when summing the contributions of a large number of base pairs. We therefore concluded that both effects result from collective motions that are facilitated by the divergence of the correlation length at denaturation.

$1/f$ fluctuations of the temperature - or equivalently of the total potential energy - have been predicted to occur in a variety of biophysical systems [19-22]. The frequency range where the $1/f$ dependence is expected to take place extends down to about 1 GHz [19-22] or even a few MHz (this work). In contrast, the recent work of Nagapriya *et al* [9] is one of the few experimental studies dealing with temperature fluctuations in such systems. According to Fig. 3 of Ref. [9], it appears that the measured power spectrum decays much more rapidly than $1/f$ in the investigated 0.01-1 Hz range. It is therefore probable that slow processes not included in our models are at work in this range. Many experimental and theoretical efforts will certainly be necessary to bridge the gap between the frequency ranges amenable, respectively, to experimental and theoretical investigations and complete our understanding of the corresponding dynamics.

# REFERENCES

[1] D. Poland and H.A. Scheraga, *Theory of Helix-Coil Transitions in Biopolymers* (Academic Press, New York, 1970)

[2] R.M. Wartell and A.S. Benight, Phys. Rep. 126, 67 (1985)

[3] M. Peyrard, Nonlinearity 17, R1 (2004)

[4] O. Gotoh, Adv. Biophys. 16, iii (1983)

[5] D. Poland and H.A. Scheraga, J. Chem. Phys. 45, 1464 (1966)

[6] E. Carlon, E. Orlandini and A.L. Stella, Phys. Rev. Lett. 88, 198101 (2002)

[7] T. Dauxois, M. Peyrard and A.R. Bishop, Phys. Rev. E 47, R44 (1993)

[8] M. Joyeux and S. Buyukdagli, Phys. Rev. E 72, 051902 (2005)

[9] K.S. Nagapriya, A.K. Raychaudhuri and D. Chatterji, Phys. Rev. Lett. 96, 038102 (2006)

[10] M. Techera, L.L. Daemen and E.W. Prohofsky, Phys. Rev. A 40, 6636 (1989)

[11] R.D. Blake, J.W. Bizzaro, J.D. Blake, G.R. Day, S.G. Delcourt, J. Knowles, K.A. Marx and J. SantaLucia, Bioinformatics, 15, 370 (1999)

[12] S. Buyukdagli, M. Sanrey and M. Joyeux, Chem. Phys. Lett. 419, 434 (2006)

[13] A. Brünger, C.B. Brooks and M. Karplus, Chem. Phys. Lett. 105, 495 (1984)

[14] T. Shimomura, K. Denda, A. Kitamura, T. Kawaguchi, M. Kito, J. Kondo, S. Kagaya, L. Qin, H. Takata, K. Miyazawa and N. Kitamura, J. Biol. Chem. 272, 6370 (1997)

[15] R. Balian, *From microphysics to macrophysics - methods and applications of statistical physics* (Springer-Verlag, Berlin, 1991)

[16] T. Geisel, A. Zacherl and G. Radons, Phys. Rev. Lett. 59, 2503 (1987)

[17] J. Barré and T. Dauxois, Europhys. Lett. 55, 164 (2001)

[18] N. Theodorakopoulos, T. Dauxois and M. Peyrard, Phys. Rev. Lett. 85, 6 (2000)

[19] S. Buyukdagli and M. Joyeux, Phys. Rev. E 73, 051910 (2006)

[20] A.R. Bizzarri and S. Cannistraro, Phys. Lett. A 236, 596 (1997)

[21] M. Takano, T. Takahashi and K. Nagayama, Phys. Rev. Lett. 80, 5691 (1998)

[22] A.R. Bizzarri and S. Cannistraro, Physica A 267, 257 (1999)

[23] P. Carlini, A.R. Bizzarri and S. Cannistraro, Physica D : Nonlinear phenomena 165, 242 (2002)

**FIGURE CAPTIONS**

**Figure 1** : (Color online) Log-log plots, at several temperatures, of the SPD of temperature fluctuations for a 2000 bp homogeneous sequence and the DPB model [7]. Melting temperature is about 354 K. For denaturated sequences, $P(f)$ is smaller than $10^{-12}$ K$^2$Hz$^{-1}$, so that the corresponding curves do not show up in the figure. Within this model, the configuration of a sequence with $N$ bp is described by $n = N$ coordinates.

**Figure 2** : (Color online) Log-log plots, at several temperatures, of the SPD of temperature fluctuations for the 2399 bp inhibitor [14] and the JB model [8]. Melting temperature is about 335 K. Within this model, the configuration of a sequence with $N$ bp is described by $n = 2N$ coordinates.

**Figure 3** : (Color online) Log-log plots, at several temperatures, of the SPD of temperature fluctuations for the 1793 bp actin (NCB entry code NM_001101) and the BSJ model [12]. Melting temperature is about 385 K. Within this model, the configuration of a sequence with $N$ bp is described by $n = 4N$ coordinates.

**Figure 4** : (Color online) Same as Fig. 2, except that SPD curves were obtained by integrating Langevin's equations of motion instead of Hamilton's ones. A dissipation coefficient $\gamma=5$ ns$^{-1}$ was assumed.

**Figure 5** : (Color online) Log-log plots, for increasing values of the dissipation coefficient $\gamma$, of the SPD of temperature fluctuations for the 2399 bp inhibitor [14] at 335 K (denaturation temperature) and the JB model [8].

**Figure 6** : (Color online) Top : plot, as a function of their position *k* in the sequence, of the average distance between paired bases for the 2399 bp inhibitor [14] at 325 and 330 K and the JB model [8]. Bottom : plot, as a function of *k*, of the AT percentage averaged over 40 consecutive base pairs of the same sequence.

**Figure 7** : (Color online) Log-log plots, at the melting temperature (335 K) and for the JB model, of the product of the SPD times the length $N_j$ of the subsection, for the 11 subsections of the 2399 bp inhibitor [14] (see Fig. 6). The thicker line denotes the curve for subsection B, which is the only closed one at the critical temperature.

**Figure 8** : (Color online) Time evolution, at 325 K and 335 K (melting temperature), of the temperature of the 2399 bp inhibitor [14] computed with the JB model [8]. Each pixel corresponds to the temperature averaged over 10 successive base pairs and 50 ns. Temperatures range from -10 K (blue/dark pixels) to +10 K (yellow/bright pixels) relative to the average one.

**Figure 9** : (Color online) Log-log plots of the SPD of temperature fluctuations for the 2399 bp inhibitor [14] at 335 K (melting temperature) and the JB model [8]. Each curve corresponds to a different value of the length $N_j$ of the subsections over which temperature fluctuations are computed.

**Figure 10** : (Color online) Evolution of the power of temperature fluctuations of the 2399 bp inhibitor [14] as a function of $N_j$ for exponentially increasing frequencies *f* (2, 20 and 200

MHz and 2 GHz) and three typical temperatures : 315 K (double stranded DNA, top plot), 335 K (melting temperature, middle plot) and 345 K (single stranded DNA, bottom plot).

**Figure 1**

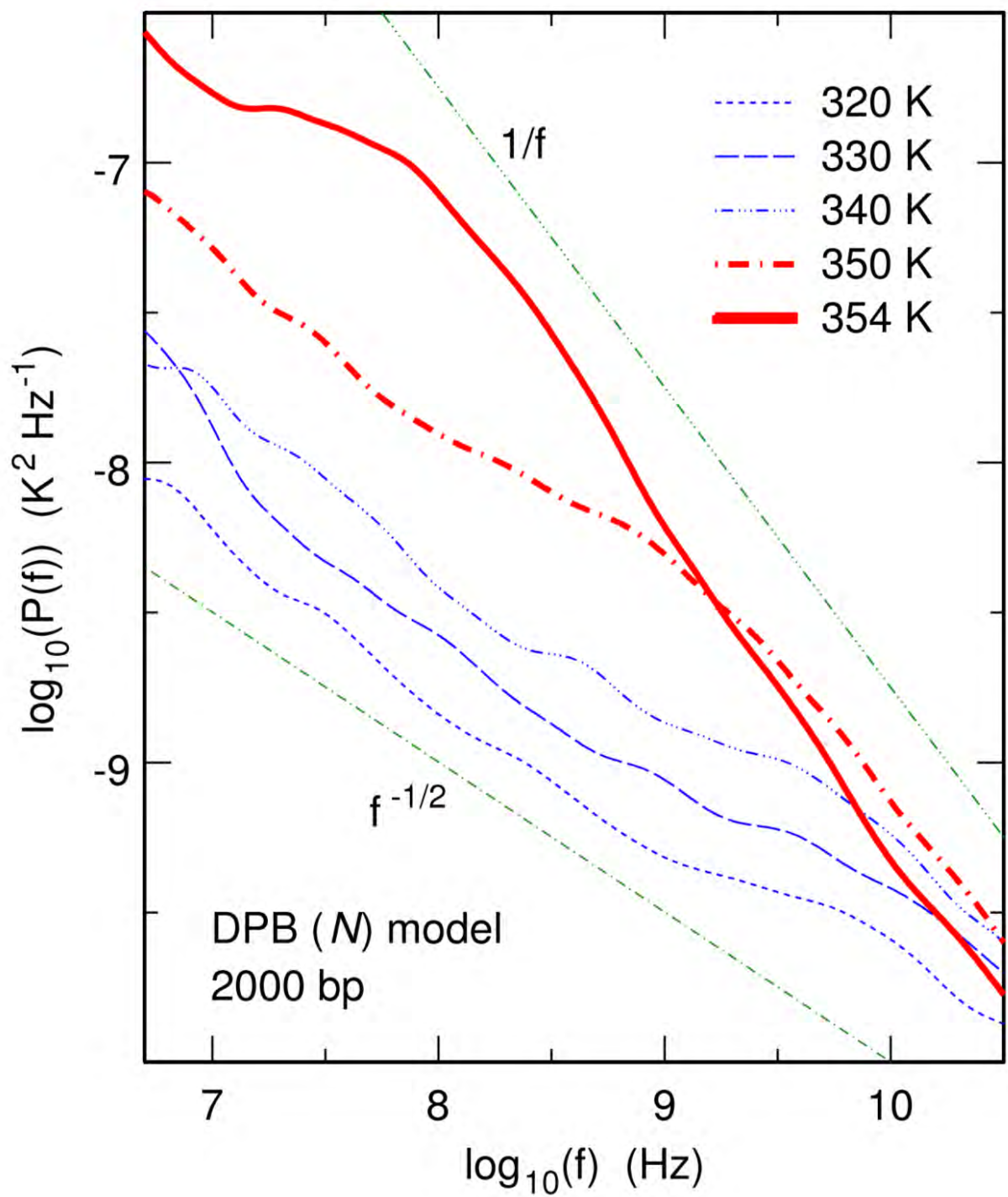

**Figure 2**

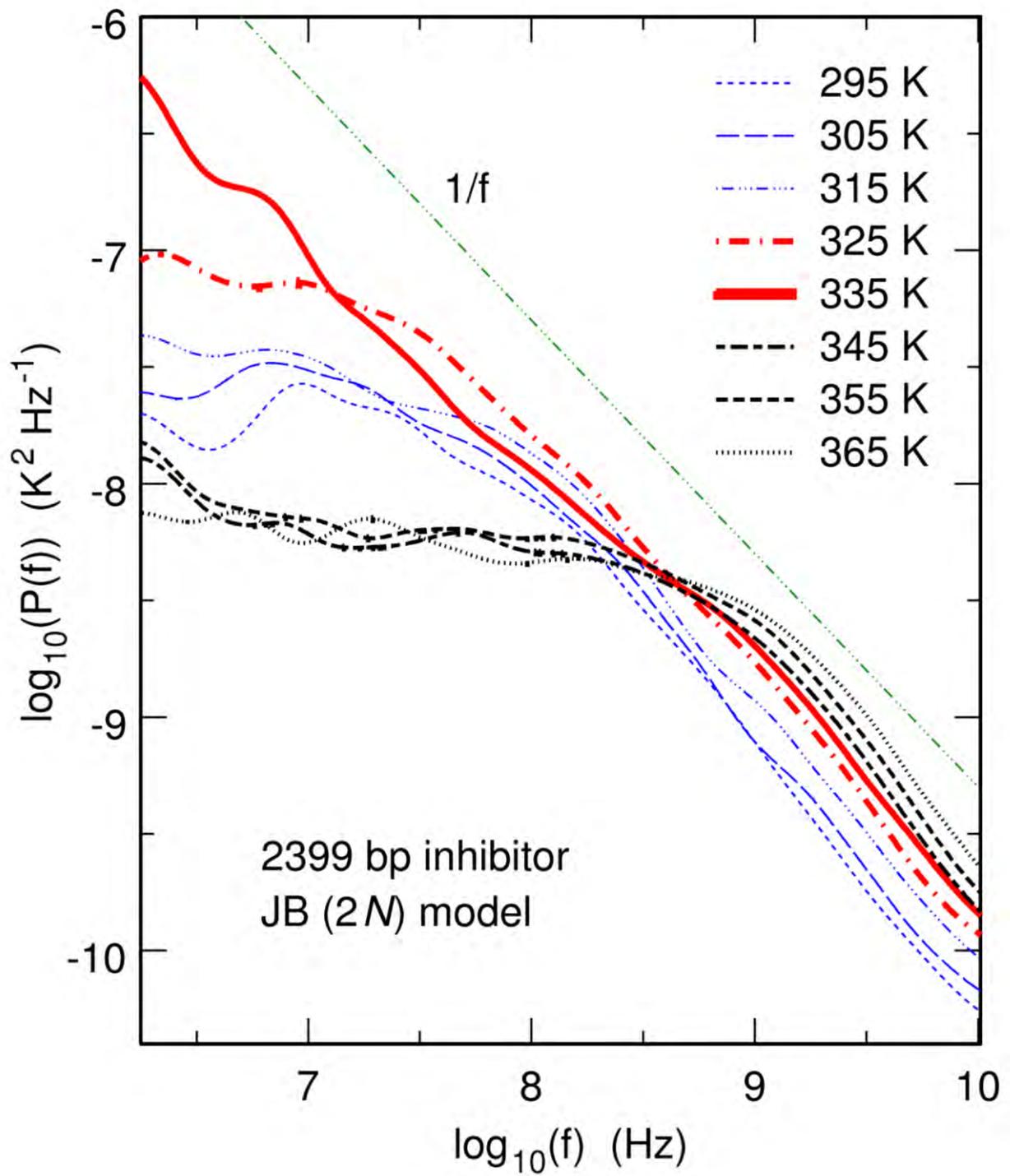

**Figure 3**

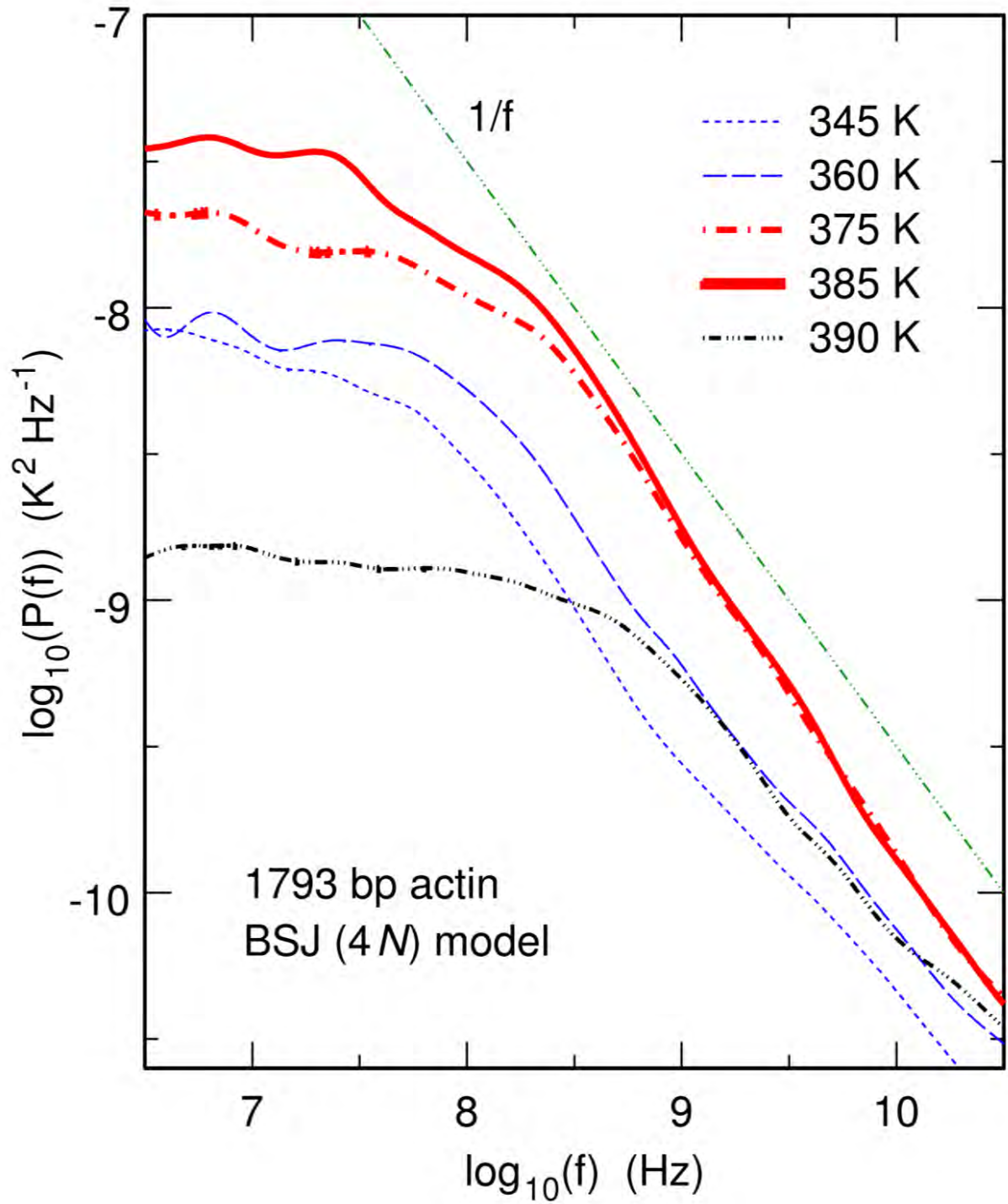

**Figure 4**

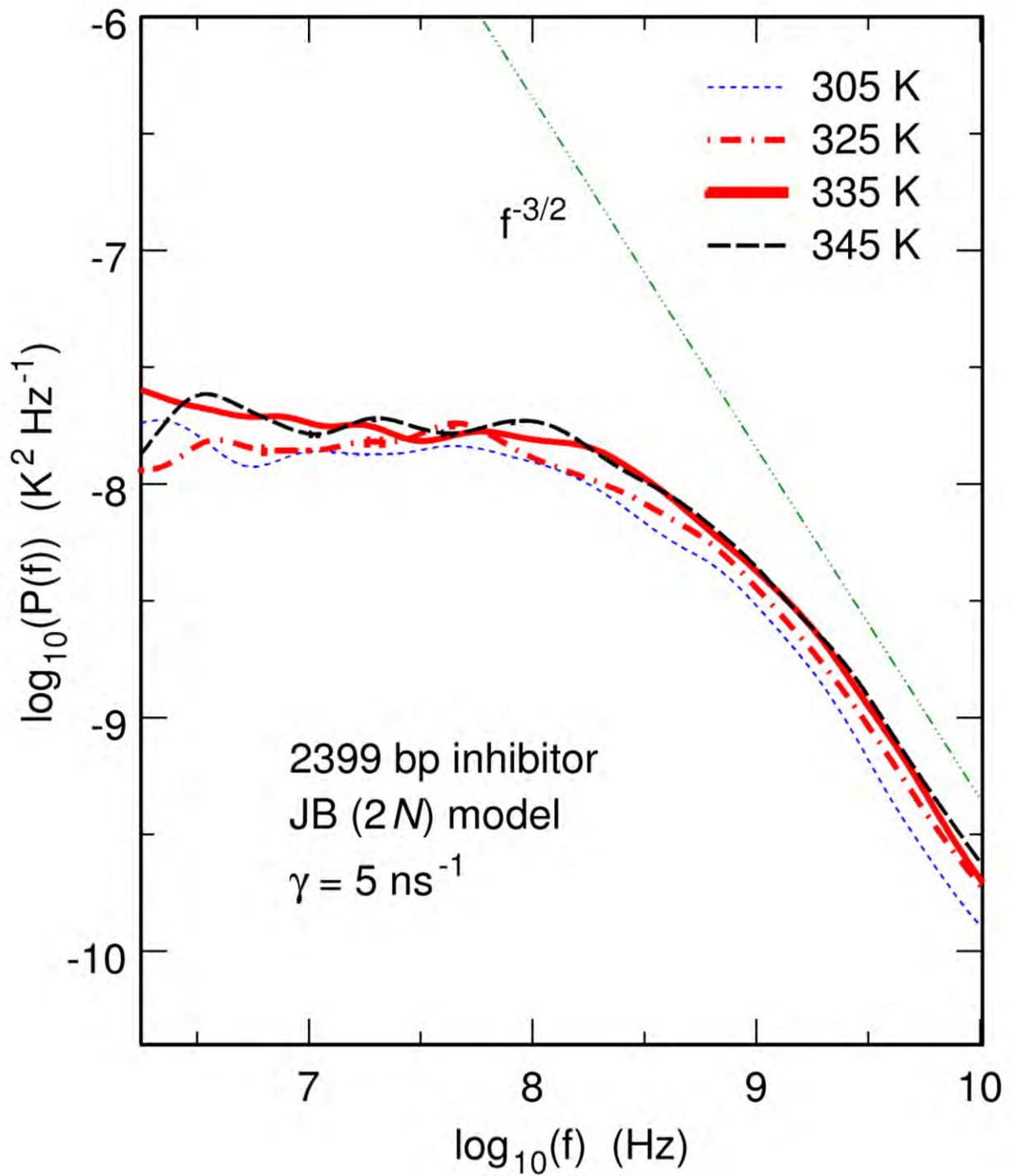

**Figure 5**

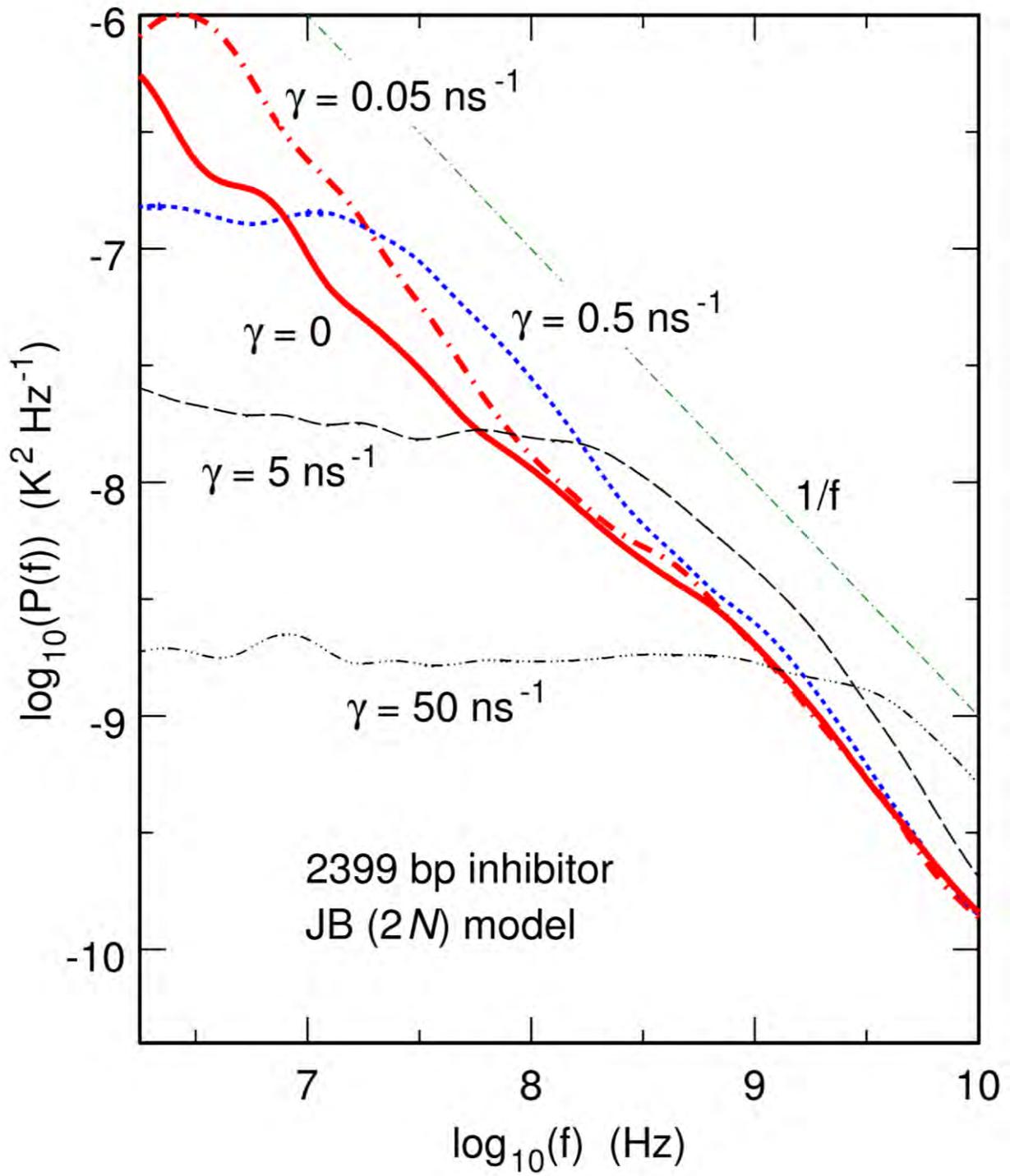

**Figure 6**

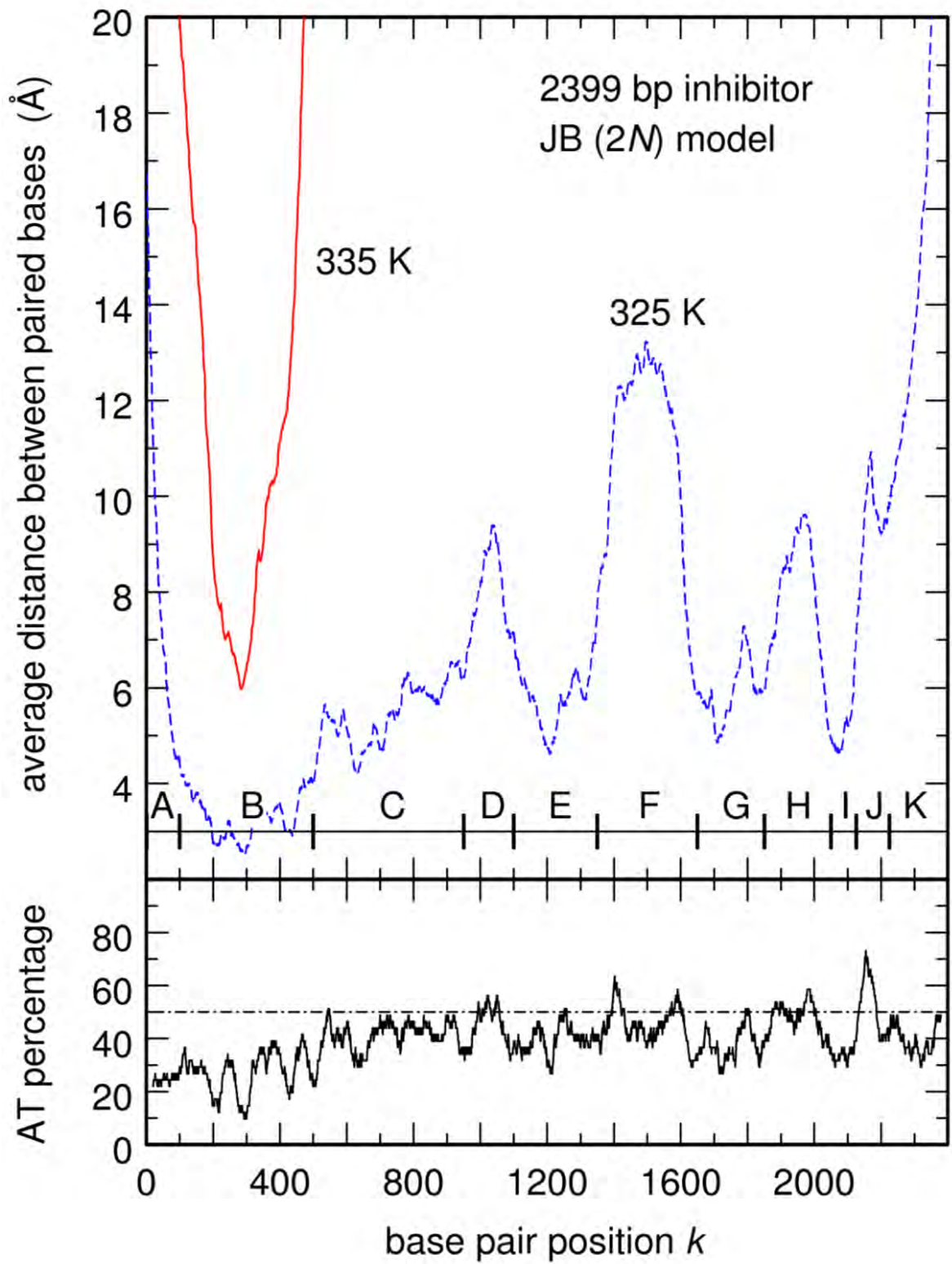

**Figure 7**

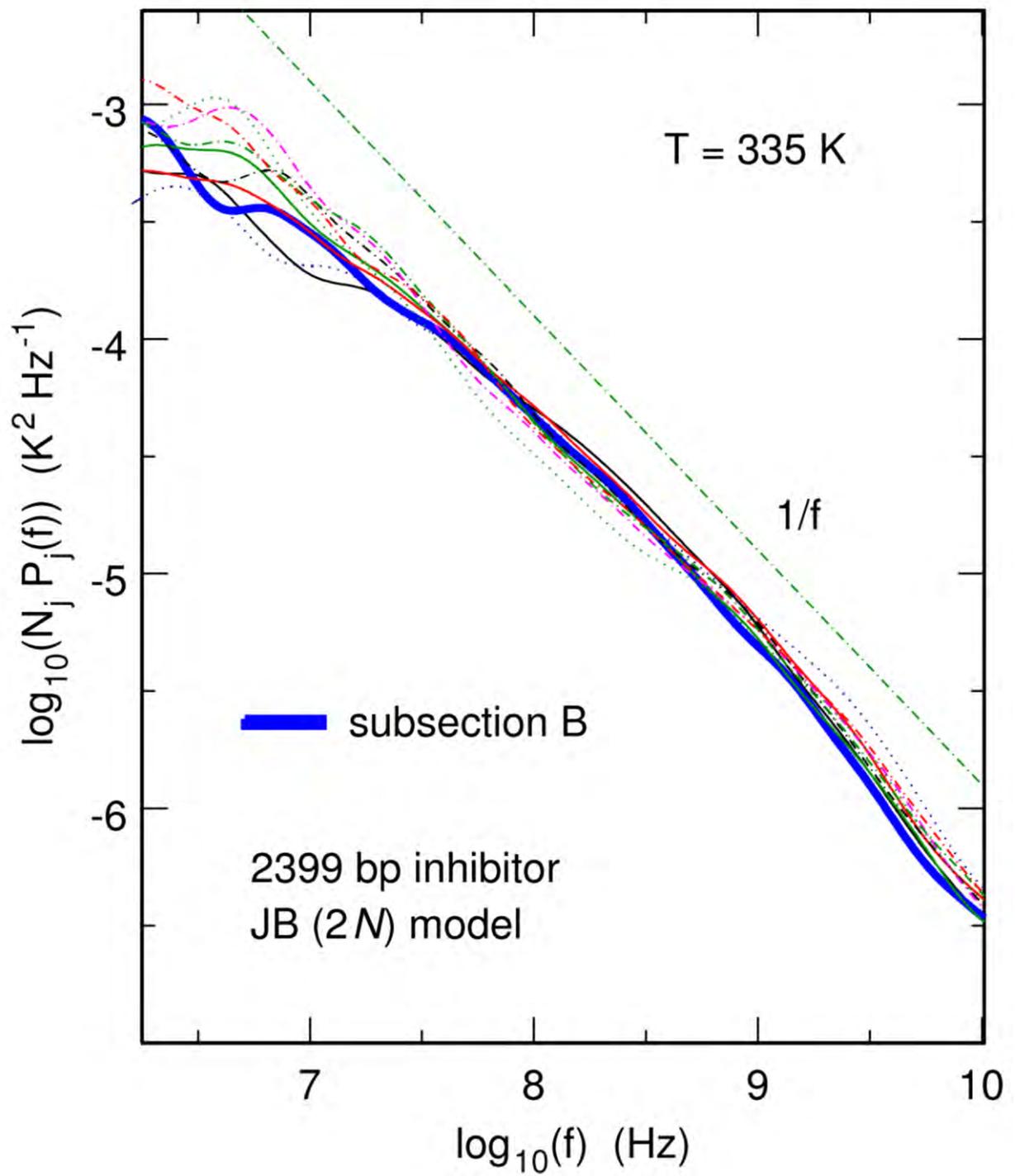

**Figure 8**

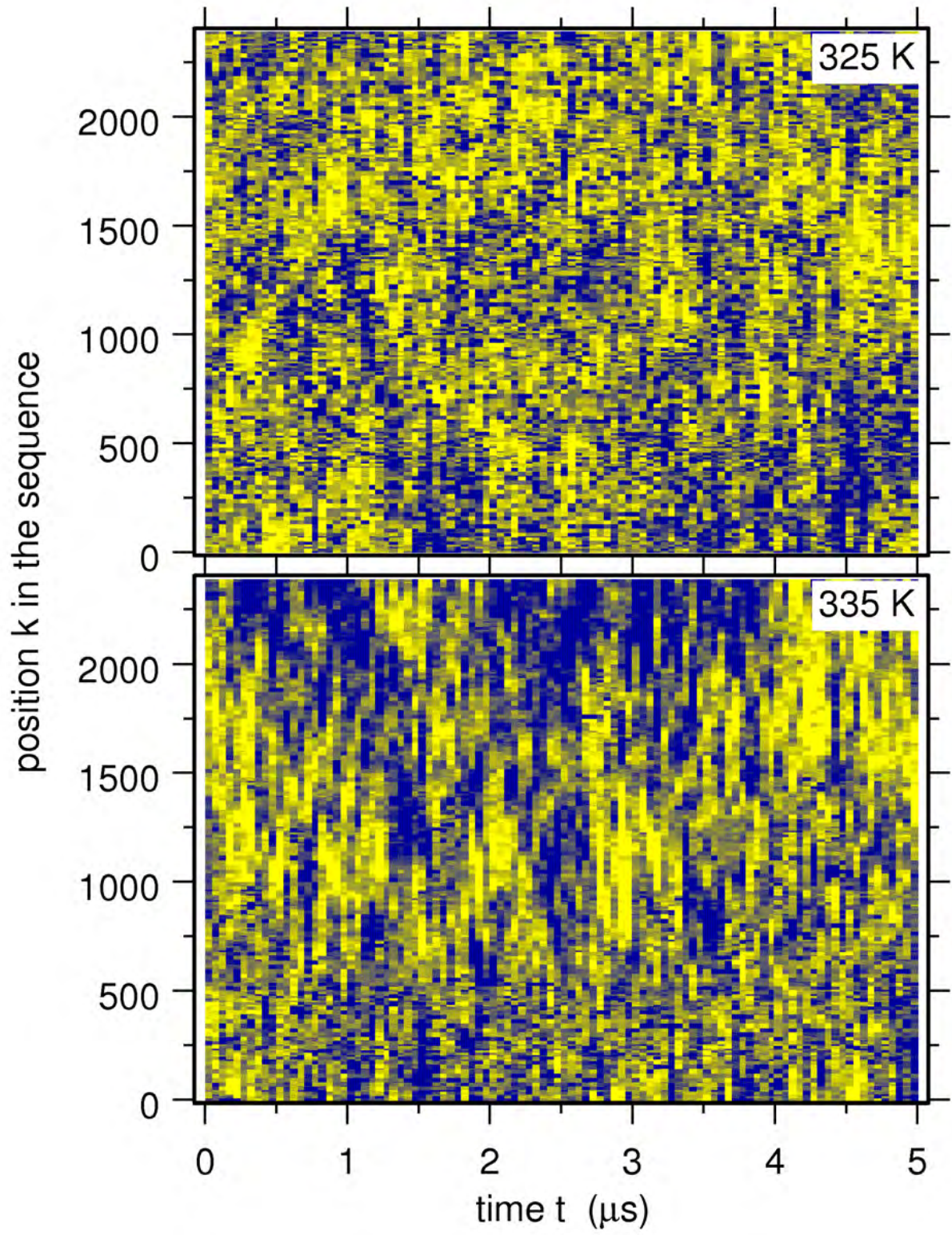

**Figure 9**

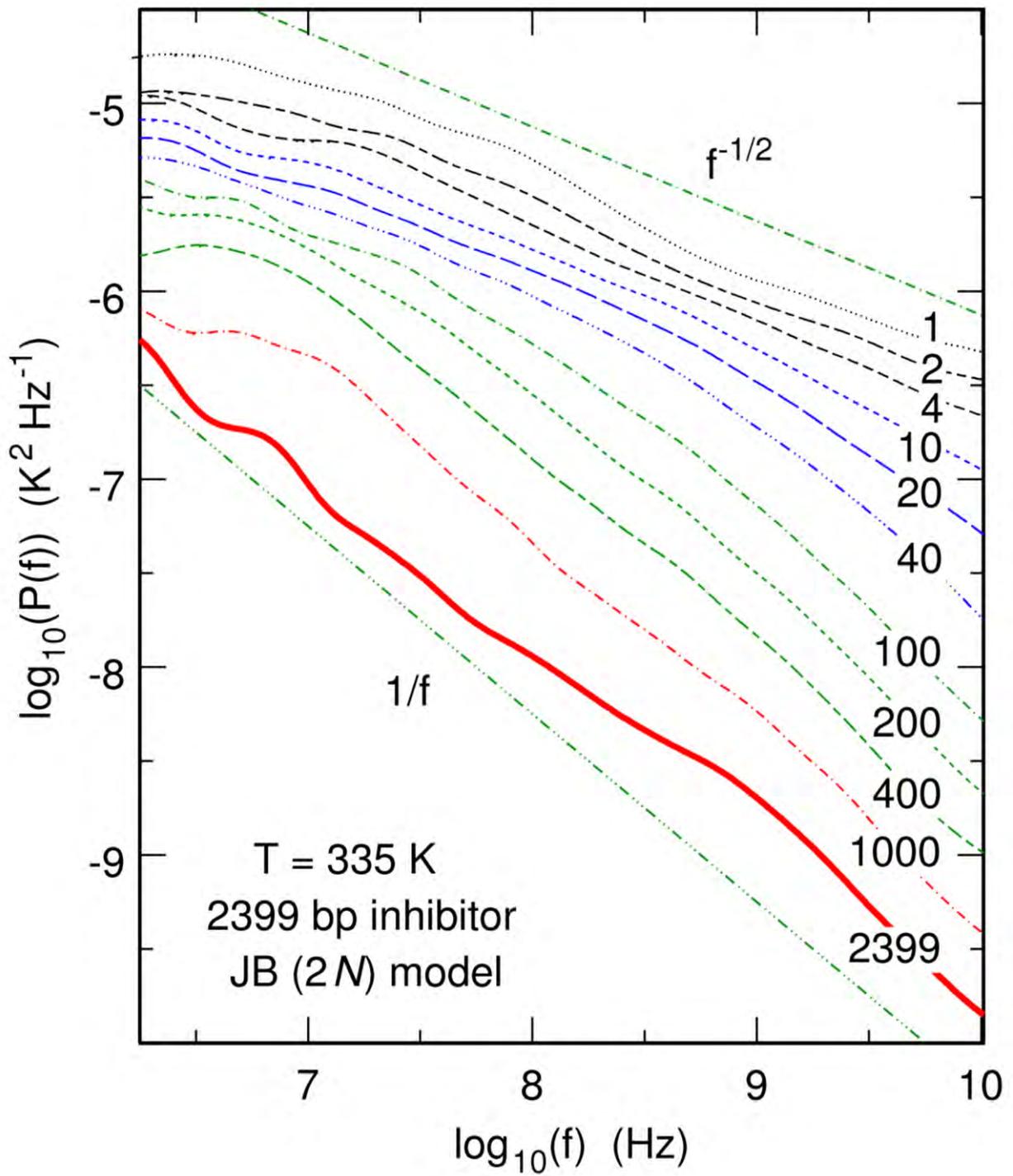

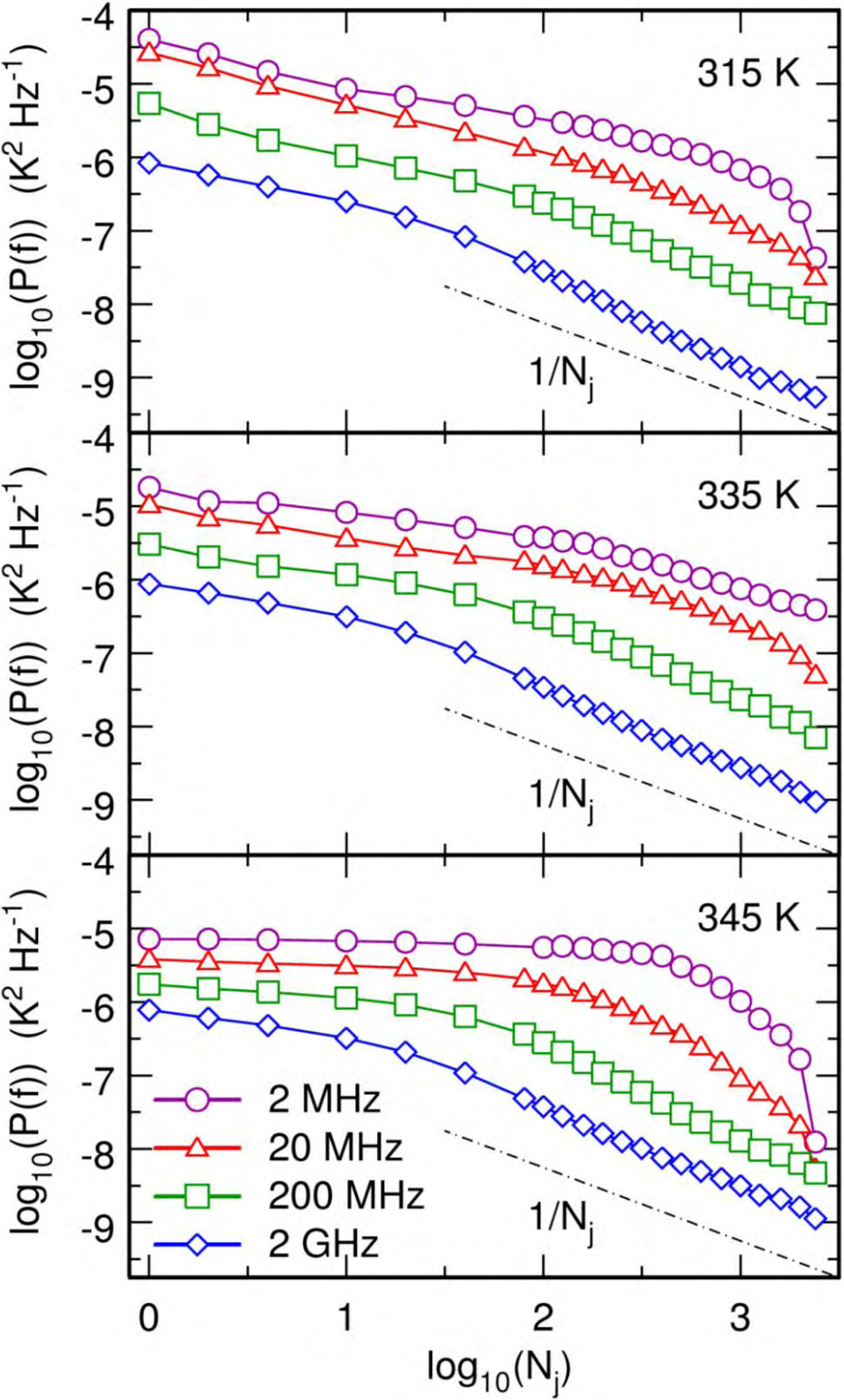